\documentclass[twocolumn,preprintnumbers,showpacs,amsmath,amssymb]{revtex4}

\usepackage{graphicx}
\usepackage{dcolumn}
\usepackage{bm}
\usepackage{textcomp}

\begin{document}

\preprint{}

\title{Lattice Relaxation in Epitaxial BaTiO$_3$ Thin Films}

\author{Feizhou He}
 \altaffiliation[Present Address: ]
 {Canadian Light Source, University of Saskatchewan, Saskatoon, SK S7N 0X4, Canada.}
\author{B. O. Wells}%
 \email{wells@phys.uconn.edu}
\affiliation{Department of Physics,
University of Connecticut, Storrs, CT 06269, USA}%

\date{\today}

\begin{abstract}
We have investigated the out-of-plane lattice relaxation related
to the ferroelectric transitions in epitaxial BaTiO$_3$ (BTO)
films using synchrotron X-ray diffraction. Under either
compressive strain or tensile strain, there is evidence
for two structural phase transitions as a function of temperature. The
transition temperature $T_C$ is a strong function of strain, which
can be as much as 100 K above the corresponding $T_C$ in bulk.
Under compressive strain, the tetragonality of BTO unit cell
implies that the polarization of the first ferroelectric phase is
out-of-plane, while under tensile strain, the polarization is
in-plane. The transitions at lower temperature may correspond to
the $aa\rightarrow r$ or $c\rightarrow r$ transitions, following
the notations by Pertsev \textit{et al}. The orientations of the
domains are consistent with theoretical predictions.
\end{abstract}

\pacs{77.55.+f, 68.55.Jk, 68.35.Rh}
\maketitle

Perovskite films have received a great deal of interest lately due
to the potential for creating working technologies based on a
variety of interesting properties such as high-$T_c$
superconductivity, colossal magneto-resistivity, ferroelectricity,
and variable dielectric constants. These properties can be quite
different in thin films versus nominally similar bulk crystals. The
primary reasons for the property changes are believed to be strain
and defects.\cite{Canedy00}

Bulk BrTiO$_3$ (BTO) undergoes three phase transitions at 393 K, 278
K and 183 K, from cubic paraelectric phase at high temperature to
three ferroelectric phases at lower temperature, with tetragonal,
orthorhombic and rhombohedral symmetries, respectively. At each
transition the lattice parameters change drastically. Thus in bulk,
the lattice parameters can be used to identify the phase
transitions. In strained BTO films, theoretical calculations based
on either first-principles method or Landau-Devonshire-type
thermodynamic theory predict that there are two successive phase
transitions for many values of film
strain.\cite{Dieguez04,Lai05,Pertsev99,Pertsev98,Ban02} The
high-temperature, paraelectric-to-ferroelectric transition has been
identified by electrical measurements.\cite{Yoneda93,Yoneda01}
However, lattice parameter measurements on epitaxial BTO films have
not been consistent. Terauchi \textit{et al} reported that both the
out-of-plane and in-plane lattice parameters increase linearly with
temperature from 15~K to 800~K, with no indications of the
transitions.\cite{Terauchi92,Yoneda93}. On the contrary, recent
experiments revealed that, the temperature dependence of the lattice
parameters do show slope changes associated with the ferroelectric
transition.\cite{Choi04,Bai04}

Experimental evidence for the transition at lower temperature has
not been reported for BTO films. This may be because this transition
involves only a slight change in the orientations of the
polarization, thus the signature is too subtle for electrical
measurements or Raman scattering. 
Inspired by the recent experimental observations and theoretical
results\cite{Alpay04} and our understanding on other perovskite film
systems,\cite{He04,He05} we have pursued an investigation of the
temperature dependence of the lattice parameters of BTO films at
lower temperatures. We show that an accurate lattice parameter
measurement does reveal a lower temperature phase transition for BTO
films. The structural evidence is consistent with the phase
transitions predicted by theory.

BTO films were grown on (001) KTaO$_3$ (KTO, $a=3.989$~\AA) and
(001) MgO ($a=4.213$~\AA) single crystal substrates by pulsed laser
deposition. The films, with thickness of about 400~\AA\ on KTO and
500~\AA\ on MgO, show excellent epitaxy with mosaics around
0.2\textdegree. X-ray diffraction measurements were carried out at
beamline X22A and X22C at the National Synchrotron Light Source,
Brookhaven National Laboratory.
The angular resolution with a graphite (002) analyzer was less
than 0.006\textdegree\ FWHM for an (0~0~2) peak, as measured from
the substrates. The temperature scans were carried out in a high
temperature capable Displex with a base temperature near 15 K and
a maximum of 800 K. The temperature control was within $\pm$0.5~K.

At room temperature, BTO in the tetragonal phase is almost
coherent with KTO substrate, with in-plane lattice parameters of
3.995~\AA\ and out-of-plane 4.041~\AA. On MgO, since the lattice
mismatch is too large, the BTO is relaxed, with in-plane lattice
parameters 4.029~\AA\ and out-of-plane 3.996~\AA. The misfit
strain is defined as $\varepsilon_m=(a_f-a_{f0})/a_f$, where $a_f$
is the measured in-plane lattice parameters, and $a_{f0}$ the
equivalent cubic cell constant of free BTO crystal. Here the
measured $a_f$ corresponds to the effective substrate lattice
parameters $b^*$ \cite{Speck94} used in theoretical calculations.
A positive or negative strain means the film is stretched in-plane
or compressed in-plane, respectively. Since the misfit strain is
calculated based on $a_{f0}$ instead of the real bulk value, even
though BTO and KTO have similar in-plane lattice parameters at
room temperature, the BTO film on KTO substrate is under a -0.28\%
(compressive) strain. The BTO film on MgO has a tensile strain of
0.57\%.

In this report, we choose a coordinate system such that
the axis normal to the film surface is the $c$ axis. The
definitions of the possible phases follow Pertsev's
notation.\cite{Pertsev98}

\begin{figure}
\includegraphics[scale=0.8]{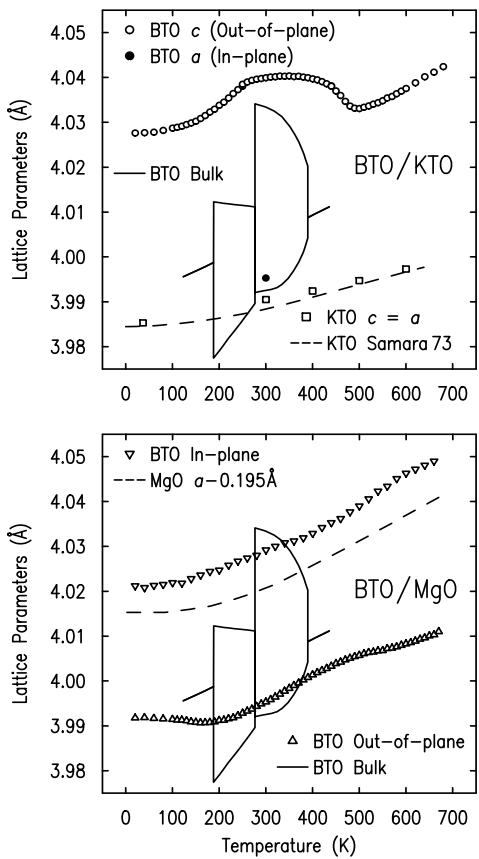}
\caption{\label{fig:bto-lat}Lattice parameters of BTO films on KTO
and MgO substrates. The curve for bulk MgO is shifted for
comparison with in-plane BTO data. Bulk BTO data from
Ref.~\cite{Kay49}. Bulk KTO data (dashed-line in upper panel) from
Ref.~\cite{Samara73}. Bulk MgO data from Ref.~\cite{White66}.}
\end{figure}

Fig.~\ref{fig:bto-lat} shows the temperature dependence of lattice
parameters for both BTO/KTO and BTO/MgO samples. We observed two
turning points in each out-of-plane lattice parameter curve. For
BTO/KTO sample, the two temperatures are at $T_1\cong500$~K and
$T_2\cong250$~K. For BTO/MgO sample, $T_1\cong450$~K and
$T_2\cong200$~K. The first turning points at higher temperature may
correspond to the paraelectric-to-ferroelectric transitions. The
transition temperature $T_1$ is much higher than the $T_C$ in BTO
bulk. The transitions at lower temperature, $T_2$, may correspond to
the $c\rightarrow r$ or $aa\rightarrow r$ transitions as predicted
by theories. As expected and shown in BTO/MgO case, the in-plane
lattice parameters of the BTO films vary smoothly over the entire
temperature range studied with no changes connected to the phase
transition. These in-plane parameters track the substrate lattice
and neither MgO nor KTO has a structural phase transition at these
temperatures.

\begin{figure}
\includegraphics[scale=0.8]{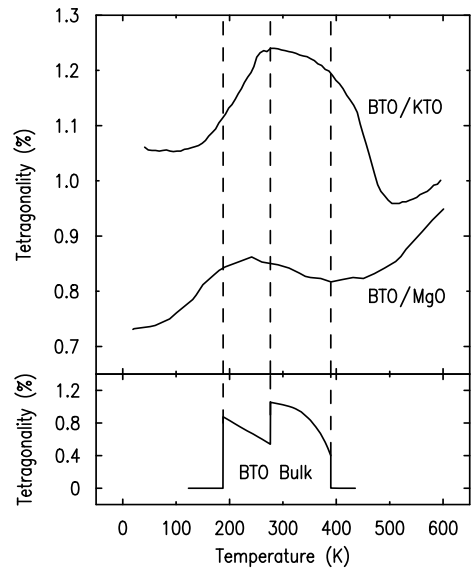}
\caption{\label{fig:bto-tetra} Tetragonality of unit cell in BTO
films suggests second order transitions. The two turning points in
both curves may indicate the onsets of polarization along
different directions. Note that the definitions of tetragonality
are slightly different for compressive strain and tensile strain
conditions. Bulk data are derived from Ref.~\onlinecite{Kay49} and
shows clearly first order transitions.}
\end{figure}

For the ferroelectric transition, the primary order parameter is the
spontaneous polarization \textbf{P}, which is not accessible by
X-ray diffraction. A secondary order parameter for the high
temperature transition is the tetragonality of the unit cell, and
that can be evaluated by X-ray diffraction. For films, in order to
maintain the convention of labelling the out of plane axis as $c$,
the definition of tetragonality is a little different for the
compressive and tensile cases. For compressive strain, the
out-of-plane $c$ is larger, so the tetragonality $\gamma=(c/a)-1$.
For tensile strain, in-plane lattice parameters are larger, thus
$\gamma=(a/c)-1$. For the lower temperature transition, and for all
transitions in films, the appropriate secondary order parameter
would be the change in slope of $\gamma$; $\gamma$ itself is not
zero above the critical temperature in any of these cases. The
temperature dependence of the tetragonality of our films are shown
in Fig.~\ref{fig:bto-tetra}. $\gamma$ from bulk BTO is shown as a
reference.

Analyzing $\gamma$ gives insight into the transition in several
ways. Particularly for BTO on MgO, the transitions are more clearly
seen. The figure makes clear that adjusted for the different
orientation of the long axis, the transitions in the two films are
similar though with different critical temperatures. Above the first
tuning point $T_1$, the tetragonality decreases smoothly. Below
$T_1$, the slopes of the out-of-plane lattice parameters change to
allow for a marked increase in $\gamma$. This increasing
tetragonality ought to reflect the internal polarization as in the
bulk. For the $c$ phase in BTO/KTO, the polarization drives $c$ axis
longer to increase tetragonality. Interestingly, under tensile
strain, although the lattice parameters cannot increase along the
polarization direction (in-plane), the BTO still manages to increase
the tetragonality by shrinking the out-of-plane lattice. Below the
$T_2$, the trend of $c$ lattice parameters changes again, so that
the tetragonality is smaller. This is consistent with the
theoretical picture that the $r$ phase is emerging.

In bulk BTO, all the phase transitions are first order, as
indicated by the appreciable discontinuities in lattice
parameters. But in epitaxial films, as illustrated by the
secondary order parameter, the phase transition becomes second
order. We see that the combination of strain and lattice
constraint imposed by epitaxy with the substrate lowers the order
of the phase transitions and broadens the transition width.

In the $c$ phase, both the polarization and the 4-fold axis of the
tetragonal unit cell are out-of-plane, so the two in-plane axes
are identical. Thus the $c$ phase is always single domain. The
tensile strain case is more complicated. The polarization has
in-plane components. If $ac$ phase exists, the $a$ axis can then
be distinguished from $b$ axis due to polarization and there may
be a difference in the length of the $a$ and $b$ axes. We would
expect twin domains with an in-plane misalignment near
90\textdegree. If $aa$ phase or $r$ phase exists, the two in-plane
axes will be identical, so the two in-plane lattice parameters
must be the same. Structurally there should be only one domain,
though internal polarization may be along different diagonals.
Thus a peak split in in-plane direction may serve as an indication
to identify what phase is present.

\begin{figure}
\includegraphics[scale=0.8]{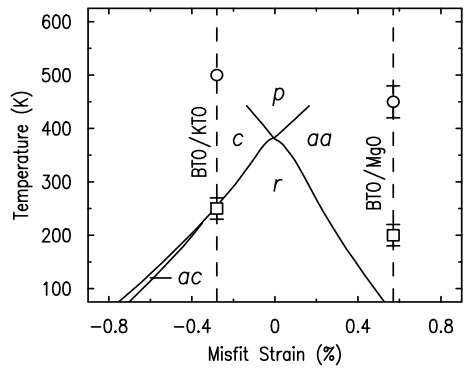}
\caption{\label{fig:bto-pd} Strain phase diagram of BTO films. The
solid lines are from theory (Ref.~\onlinecite{Pertsev99}) by
Pertsev.}
\end{figure}

By using reciprocal spacing mapping, we examined the domain
structure in BTO/MgO sample. Through the whole temperature range
we probed, there is no peak splitting observed. This indicates
that the $aa$ phase is the likely state between $T_1$ and $T_2$.
Note that this $aa$ phase is slightly different from the
orthorhombic phase found in bulk. In bulk, the pseudo-cubic unit
cell for orthorhombic phase is elongated along one face diagonal
direction, which is the natural consequence of the polarization.
The $aa$ phase, however, has a square in-plane lattice due to the
substrate constraint.

This result supports the predictions in
Ref.~\cite{Dieguez04,Lai05,Pertsev99} that $p$, $c$, $aa$ and $r$
phases are presented while $ac$ phase is unlikely. As an example, in
Fig.~\ref{fig:bto-pd} we compare our data with the strain phase
diagram calculated by Pertsev \textit{et al}.\cite{Pertsev99} While
we have only a limited number of experimental data on this phase
diagram, several points are brought to light. The general layout of
the phase diagram, including both the specifically predicted phases
and the rough critical temperatures, is consistent with our
experimental observations. While the $T_C$s agree fairly well with
theory in a qualitative sense, quantitative agreement is present
only for compressive strain. For tensile strain, our data show a
substantially smaller change in $T_C$ versus the unstrained case
than predicted in Ref.~\cite{Pertsev99}. The calculation of Lai
\textit{et al}. in Ref.~\cite{Lai05} shows better agreement with the
data for positive strain. Another phase diagram obtained through
first-principles method shares the same topology despite a shift in
temperatures.\cite{Dieguez04} However, we note that there is a large
lattice mismatch between BTO and MgO and these films are not
coherently grown.

There are some aspects of the predicted phase diagram that we cannot
yet corroborate. We do not have enough data to verify that a
tetra-critical point, sometimes called a phase point, occurs near
zero strain. It may be that the smaller than predicted change in
spread of critical temperatures for tensile strain indicates that
this phase point is either at larger strain values, as suggested in
Ref.~\cite{Lai05}, or that the structure of the phase diagram near
zero strain is more complicated than predicted. These questions
require further study.

In summary, epitaxial BTO films were grown on KTO and MgO
substrates to induce compressive strain and tensile strain,
respectively. Through temperature dependence of the lattice
parameters, two phase transitions can be identified in each
sample. The tetragonality analysis implies that the orientations
of the spontaneous polarization are consistent with theoretical
predictions.

We acknowledge S. P. Alpay and B. Misirlioglu for helpful
discussion. This material is based upon work supported by the
National Science Foundation under Grant No. DMR-0239667. BW thanks
the Cottrell Scholar Program of the research corporation for
partial support of this work. Work at Brookhaven is supported by
Division of Material Sciences, U.S. Department of Energy under
contract DE-AC02-98CH10886.


\end{document}